\documentclass[twocolumn]{aastex631}

\accepted{February 24, 2026}
\submitjournal{ApJL}

\shortauthors{Calamari et al.}

\usepackage{amsmath}
\usepackage{mathrsfs}
\usepackage{natbib}
\usepackage{xcolor}
\usepackage{threeparttable}
\usepackage[hang,flushmargin]{footmisc}

\begin{document}

\title{Bridging the Gap: Using Brown Dwarfs to Examine Silicate Clouds in Giant Exoplanet Atmospheres}

\author[0000-0002-2682-0790]{Emily Calamari}
\affiliation{Department of Astrophysics, American Museum of Natural History, New York, NY 10024, USA}

\author[0000-0001-6251-0573]{Jacqueline K. Faherty}
\affiliation{Department of Astrophysics, American Museum of Natural History, New York, NY 10024, USA}

\author[0000-0001-6627-6067]{Channon Visscher}
\affiliation{Department of Chemistry and Planetary Sciences, Dordt University, Sioux Center, IA, USA}
\affiliation{Center for Exoplanetary Systems, Space Science Institute, Boulder, CO, USA}

\author[0000-0002-8871-773X]{Marina E. Gemma}
\affiliation{Department of Astrophysics, American Museum of Natural History, New York, NY 10024, USA}
\affiliation{Department of Earth and Planetary Sciences, American Museum of Natural History, New York, NY 10024, USA}

\author[0000-0003-4083-9962]{Austin Rothermich}
\affiliation{The Graduate Center, City University of New York, New York, NY 10016, USA}
\affiliation{Department of Astrophysics, American Museum of Natural History, New York, NY 10024, USA}

\author[0000-0003-0150-3489]{Francisco Ardévol Martínez}
\affiliation{Department of Astrophysics, American Museum of Natural History, New York, NY 10024, USA}

\author[0000-0003-0548-0093]{Sherelyn Alejandro Merchan}
\affiliation{The Graduate Center, City University of New York, New York, NY 10016, USA}
\affiliation{Department of Astrophysics, American Museum of Natural History, New York, NY 10024, USA}

\author[0000-0002-2011-4924]{Genaro Suárez}
\affiliation{Department of Astrophysics, American Museum of Natural History, New York, NY 10024, USA}

\begin{abstract}
We present results from examining the silicate cloud modeling of four JWST-observed hot Jupiters in the context of brown dwarf theory to further explore signatures of formation in present-day atmospheres. We contextualize our understanding of protoplanetary disk refractory chemistry with empirical evidence from chondritic meteorites to show that giant planets forming and accreting in the outer disk adopt their stellar Mg/Si value. We show that current silicate cloud species determinations of WASP-17 b, WASP-107 b, WASP-39 b and HD 189733 b are in line with predictions laid out in \citet{Calamari2024} based on each system's host star Mg/Si ratio, further supporting this hypothesis. We discuss physical motivations for potential atmospheric scenarios where apparent silicate cloud species is not in agreement with that predicted by its host star chemistry. Additionally, we compare current transit spectroscopy for three of these four exoplanets against brown dwarf spectra to examine molecular absorption trends across the substellar mass temperature regime.
\end{abstract}

\section{Introduction}\label{sec:intro}
Since the very first endeavors to model brown dwarf spectra, their similarities to the gas giant planets were clear. Some initial attempts to model the spectrum of the first discovered methane-bearing brown dwarf, Gliese 229 B, adapted Jovian models into the brown dwarf temperature regime, noting its likeness to Jupiter \citep[e.g.][]{Marley1996Sci, Burrows1997, Burrows1999}. While formation pathway may be used as a distinguishing factor between brown dwarfs and giant exoplanets (as it certainly does create differences in metallicity, rotation rate, etc.), their overlapping temperatures, masses, radii and ages have resulted in unified atmospheric and evolutionary theory \citep[][etc.]{Burrows1997, Marley1997, burrowssharp1999, Burrows2001, LodFeg2002, Baraffe2003, Freedman2008, Marley2015}. As a result, the ways in which we approach atmospheric modeling (deriving temperature-pressure (T-P) profiles and bulk chemical makeup, determining cloud presence and species, etc.) overlap significantly. As exoplanet science has seen major advancements in the resolution and quality of their datasets as a result of the James Webb Space Telescope (JWST), so, too, has brown dwarf science. In order to best understand and interpret giant exoplanet spectra, it is important we apply the lessons learned and advancements made in modeling brown dwarf atmospheres.

Several studies have tried to elucidate the connection between the measured chemical abundances of planet hosting stars and their companions \citep[e.g.,][and references therein]{Gonzalez1998, Fischer2005, Teske2014, Teske2019, Wilson2022, Teske2024} with the larger goal of understanding how and why planetesimals begin to form. Studies looking at global chemical parameters, like total stellar metallicity, have been able to draw links to planet occurrence rate -- namely, the Planet-Metallicity Connection (PMC), which postulates that the likelihood of planetesimal growth is greater around high metallicity stars \citep{Fischer2005}. However, studies examining individual planet-hosting stellar abundances have relatively inconclusive results. \citet{Adibekyan2012} suggested enhanced magnesium (Mg) inventory in a star, and subsequent proto-planetary disk, could lead to increased probability of planetesimal growth while \citet{Teske2014} found no correlation between any given elemental abundance and planet occurrence rate. However, nearly all studies note the relatively small sample size of planet-hosting stars with individual measured chemical abundances.

While examining bulk metallicity or individual elemental abundances may not provide certainty on planet formation pathways, it may be useful in tracing an exoplanet's formation history (i.e., PMC seems to support the core accretion method of planet formation). Most notably, measuring a C/O ratio in an extrasolar world has been theorized as a key metric to tracing formation location in a protoplanetary disk. Disagreements between observed companion and stellar C/O ratio can give clues to formation location depending on a given stellar system's ice lines (locations in the disk where Mg-silicates, H$_2$O, CO, CO$_2$ gas condense out) \citep{Oberg2011, Madhusudhan2012}. Giant exoplanets with increased C/O ratios respective to their host star would support the idea that they formed and accreted material from the outer disk, where the disk gas-phase C/O ratio would be elevated relative to its stellar value. However, studies of protoplanetary disks with the Atacama Large (sub)Millimeter Array (ALMA) have shown that these theoretical predictions may not line up as seamlessly with observations, as disk chemistry proves to be unsurprisingly much more complex \citep{Oberg2021, Bosman2021a, Bosman2021b}. Connecting observable C/O ratios in exoplanet atmospheres back to disk formation location may in fact be a far off goal, especially as observable C/O ratios are likely not reflective of bulk C/O ratio in an object in large part due to cloud formation trapping a significant percentage of gaseous oxygen into silicate grains. \citep{burrowssharp1999, Line2015, Burningham2017, Calamari2022, Calamari2024}. If we want to connect the chemistry of the stellar host and planet companion, as has been done in the brown dwarf regime \citep[e.g.][]{Line2017, Kitzmann2020, Zhang2022, Calamari2024, Zhang2025}, we instead move from examining volatile (C, O) to refractory elements (Mg, Si) found in silicate clouds \citep[see also discussions in][]{Feinstein2025}.

In brown dwarf science, substellar mass objects with empirically determined fundamental parameters (i.e. luminosity via parallax, age via known stellar host star, etc.) have been designated as “benchmark" objects \citep{Pinfield2006, Faherty2010, Deacon2014}. Benchmarks are considered essential tools in breaking the age-mass-temperature degeneracy and informing our atmospheric and evolutionary models \citep[e.g.][]{Bowler2009, Burningham2009}. Recently, \citet{Calamari2024} showed the power of host star chemistry in understanding the atmospheres of companion brown dwarfs. By placing substellar mass objects in context within their larger co-moving system, \citet{Calamari2024} was able to place constraints on silicate cloud formation. Specifically, they determined how much oxygen will be sequestered in these cloud layers and what types of silicates we should expect to see based on the given host star chemistry. Despite the added complexities of many giant exoplanet atmospheres (photochemistry, differentiated interiors, tidal locking, etc.) and the asynchronous formation history of star and planet, we build off of this unified atmospheric theory in a similar attempt to link host star chemistry with their respective companion giant exoplanet atmosphere. If brown dwarf atmospheres can be considered a first order case for the giant exoplanets, we aim to find the connecting thread that allows us to expand our understanding of silicate clouds in giant exoplanets.

In this paper, we turn to recently published works on gas giant exoplanets observed by JWST where the host star chemistry is known. Because of the complexity of giant planet formation and the many variables that can contribute to observed chemical abundances (i.e. migration through a chemically non-uniform disk, various accretion scenarios, planetary interior differentiation, photochemical production/dissociation, etc.), we do not have the simplicity of assuming co-evality between a host star and its companion as we do in brown dwarf companion systems. As a result, we cannot directly compare chemical compositions between a planet-hosting star and a companion exoplanet atmosphere to disentangle the effects of clouds on element abundance ratios and bulk chemical abundance \citep{Calamari2024}. Instead, we attempt to bridge the gap between brown dwarf and exoplanet atmospheric analysis by examining the cloud modeling of the JWST-observed exoplanet population to see if there are trailing signatures of host chemistry in planetary companions. In particular, we examine the modeled atmospheres of WASP-17 b \citep{Grant2023}, WASP-107 b \citep{Dyrek2024}, WASP-39 b \citep{Powell2024} and HD 189733 b \citep{Inglis2024}.


In \autoref{sec:systems} we outline the JWST observations and subsequent modeling of each of the four systems we will examine in detail. In \autoref{sec:host chem}, we discuss how host star chemistry can inform our understanding of exoplanet atmospheres by looking within the Solar system and also outwardly to protoplanetary disk observations. In \autoref{sec:silicates}, we place exoplanet transit spectra in context with brown dwarf spectra to better understand the observable similarities and differences of these two populations. Finally, in \autoref{sec:clouds}, we discuss the implications of current silicate clouds in modeled exoplanet atmospheres and our predictions for future cloud modeling with JWST.

\begin{deluxetable*}{lccccccccc}
\tablenum{1}
\tablecaption{Select System Parameters for JWST-Observed Hot Jupiters}
\tablehead{\colhead{\textbf{Planet}} & \colhead{\shortstack{\textbf{Stellar}\\ \textbf{SpT}}} & \colhead{\shortstack{\textbf{Stellar}\\ \textbf{Mg/Si}}} & \colhead{\shortstack{\textbf{Stellar}\\ \textbf{C/O}}} & \colhead{\textbf{Age (Gyr)}} & \colhead{\textbf{T$_{\rm Eq}$ (K)}} & \colhead{\textbf{M$_{Jup}$}}  & \colhead{\textbf{R$_{Jup}$}} & \colhead{\textbf{$a$ (au)}} & \colhead{\textbf{Cloud Model}}}
\startdata
    \hline
     WASP-17 b & F6 & 0.90$\pm0.03$ & 0.18$\pm0.02$ & 3.0$^{+0.9}_{-2.6}$ & 1771 & 0.477 & 1.932 & 0.0515 & SiO$_2$ (c)\\
     HD 189733 b & K2V & 0.90$\pm0.03$ & 0.68$\pm0.07$ & 4-8 & 1200 & 1.13 & 1.13 & 0.03142 & SiO$_2$ (a)\\
     WASP-39 b & G8 & 1.05$\pm0.03$ & 0.46$\pm0.05$ & 8.5-9 & 1116 & 0.28 & 1.27 & 0.0486 & Grey$^1$\\
     WASP-107 b & K6 & 1.08$\pm0.18$ & 0.50$\pm0.1$ & 3.4 $\pm$0.3 & 740 & 0.12 & 0.94 & 0.055 & Sil. Mixture$^2$ (a)\\
    \hline
\enddata
    \begin{tablenotes}
        \item Denotation of \textit{(c)} next to cloud species indicates the crystalline form of this condensate whereas \textit{(a)} indicates amorphous. \item References for all values in this table are laid out in \autoref{sec:systems}.
        \item $^1$Further cloud modeling is required to best understand the presence of clouds in this atmosphere.
        \item $^2$A mixture of amorphous Si-based condensate refractive indices (MgSiO$_3$, SiO, SiO$_2$) were used to replicate the observed silicate feature. See \autoref{sec:systems}.
    \end{tablenotes}
    \label{tab:systems}
\end{deluxetable*}

\section{Relevant JWST-Observed Planetary Systems}\label{sec:systems}

In this section, we review retrieval cloud modeling \citep[see][etc.]{Madu2009, Line2015, Burningham2017} of JWST MIRI transmission or emission spectra for the giant exoplanets in each of the following four systems listed in \autoref{tab:systems}. For a more detailed discussion of JWST observations or pre-JWST modeling efforts, please refer to \citet{Grant2023, Dyrek2024, Powell2024, Inglis2024} and references therein. Host star chemical abundances and abundance ratios for all systems discussed in this paper are compiled from \citet{Brewer2016, Hejazi2023}. Information in this section is also listed in \autoref{tab:systems} for ease of reading.

\subsection{WASP-17 b}
\citet{Grant2023} reports on the analysis of JWST MIRI Low Resolution Spectrometer (LRS) transit data (5-12 $\mu$m) of WASP-17 b, a hot Jupiter orbiting an F6 type host star. They cite a mass of 0.477 M$_{Jup}$, radius of 1.932 R$_{Jup}$ and equilibrium temperature (T$_{\rm eq}$) of 1771 K for WASP-17 b. WASP-17 b also has a semi-major axis of 0.0515 au \citep{Anderson2010}. Radiative-convective thermochemical equilibrium (RCTE) atmospheric models (or, forward models) were computed using a combination of PICASO v3.1  \citep{Batalha2019, Mukherjee2023} and Virga \citep{Batalha2020, Rooney2022, Batalha2025a, Batalha2025b} and fit to the combined JWST MIRI LRS data along with Hubble Space Telescope (HST) and Spitzer data presented in \citet{Alderson2022} (0.3 - 5 $\mu$m). From their forward model fits, they found that a crystalline SiO$_2$(s) (quartz\footnote{Recent work has suggested different polymorphs of crystalline silica may be more representative of the clouds in these atmospheres \citep[see][]{Moran2024}.}) cloud model was statistically preferred over other cloud species, including enstatite (MgSiO$_3$) and forsterite (Mg$_2$SiO$_4$), as well as a cloudless model.
Free chemistry retrievals were done on the combined dataset of HST + Spitzer + JWST using the retrieval code POSEIDON \citep{MacDonald2023} adapted to account for Mie-scattering aerosols following the approach in \citet{Zhang2019}. For the retrieval modeling, they similarly found that a crystalline SiO$_2$(s) cloud model was statistically preferred over both cloud-free and generic (“grey") aerosols. Finally, they validate their retrieval approach with POSEIDON by conducting retrievals with \textsc{petitradtrans} \citep[pRT;][]{Molliere2019} and again find the same statistical preference for quartz clouds. For the host star in this system, the C/O ratio is 0.18$\pm0.02$ and the Mg/Si ratio is 0.90$\pm0.03$.

\subsection{HD 189733 b}
\citet{Inglis2024} presents both free chemistry retrievals with pRT and RCTE forward model fits with PICASO + Virga (as above) on the JWST MIRI LRS emission spectrum (5-12 $\mu$m) of HD 189733 b, a hot Jupiter orbiting a K2V type star. HD 189733 b has a mass of 1.13 M$_{Jup}$, a semi-major axis of 0.03142 au \citep{Southworth2012}, a radius of 1.13 R$_{Jup}$ \citep{Sing2016} and T$_{\rm eq}$ of 1200 K \citep{Stassun2017}. Several cloud models were tested, including all three major silicate species (Mg$_2$SiO$_4$, MgSiO$_3$, SiO$_2$), and statistical evidence showed an amorphous SiO$_2$(s) cloud model to best reproduce the absorption feature at 8.7 $\mu$m and best fit the data. 
For the forward modeling, \citet{Inglis2024} shows that a quartz cloud model is strongly preferred over the cloud-free model. The authors do note that at the temperatures and pressures expected on the dayside of HD 189733 b, vertical mixing of condensate particles from higher pressures to the visible photosphere ($\sim$10$^{-2}$ - 10$^{-3}$ bar) would have to play a role in SiO$_2$(s) absorption in this spectrum. For the host star in this system, the C/O ratio is 0.68$\pm0.07$ and the Mg/Si ratio is 0.90$\pm0.03$.

\subsection{WASP-39 b}
\citet{Powell2024} presented extensive modeling results on the JWST MIRI LRS transit spectrum of WASP-39 b, a Saturn-mass exoplanet orbiting a G8 type host. WASP-39 b has a mass of 0.28 M$_{Jup}$, a radius of 1.27 R$_{Jup}$, a T$_{\rm eq}$ of 1116 K and a separation of 0.0486 au \citep{Faedi2011}. The host star for this system has a C/O ratio of 0.46$\pm0.05$ and an Mg/Si ratio of 1.05$\pm0.03$. A suite of forward models that account for photochemistry were used to analyze the gaseous content of the atmosphere as well as RCTE modeling done with PICASO + Virga to understand the impact of clouds. Cloud species modeling focused on enstatite as well as Na$_2$S and MnS. Free chemistry retrievals were also conducted using seven different retrieval codes: ARCiS \citep{Ormel2019, Min2020}, 
Aurora \citep{Welbanks2021},  
CHIMERA \citep{Line2015}, 
Helios-R2 \citep{Kitzmann2020}, 
NEMESIS \citep{Irwin2008}, 
Pyrat Bay \citep{Cubillos2021}, 
and TauREx \citep{Waldmann2015, Waldmann2015b}. 
It is important to note that the bulk of the comparative modeling focused on detecting the gaseous species SO$_2$ in this atmosphere. Additionally, the apparent decrease in transit depth beyond 10 $\mu$m complicated results as it remains unexplained. Despite the inability of Mie-scattering cloud models to detect spectral signatures in the transit data, \citet{Powell2024} do require a grey cloud prescription for each final retrieval model. The cloud parameters are poorly constrained and the authors do note that more extensive cloud modeling is required in future work in order to best estimate condensate species, cloud top pressure, cloud fraction, etc.

More recent attempts at constraining this cloud opacity in WASP-39 b have been presented in \citet{Arfaux2024} and \citet{Ma2025} with the former finding reasonable fits to near-infrared data only using Na$_2$S and enstatite clouds and the later finding that quartz and enstatite clouds improve their fit in the mid-infrared. It is likely that additional mid-infrared data at higher resolution than provided by MIRI LRS is needed in order to uncover the cloud processes at work here.

\subsection{WASP-107 b}
\citet{Dyrek2024} presents JWST MIRI LRS transit spectral analysis on the warm Neptune, WASP-107 b. WASP-107 b has a mass of 0.12 M$_{Jup}$, a radius of 0.94 R$_{Jup}$, a T$_{\rm eq}$ of 740 K and a semi-major axis of 0.055 au orbiting a K6 type star \citep{Anderson2017}. They performed two sets of atmospheric retrievals with pRT and ARCiS using a combination of JWST MIRI LRS and near-infrared (1.121-1.629 $\mu$m) HST data. They present a resulting best fit model that includes a high-altitude ($\sim$ 10$^{-5}$ bar) silicate cloud layer.
While they try a variety of cloud species, they report a best fit that includes a mixture of refractive indices for enstatite, quartz and SiO(s). However, they note that while SiO could theoretically condense at the temperature and pressure of the cloud layer, it would likely have rained out to hotter, deeper atmospheric layers. Additionally, there are questions on the plausibility of a silicate cloud layer at such a high altitude in a cool object. For this solution to be physically possible, it would require uncomfortably strong vertical mixing (i.e., $\log$${K_{zz}}$$\sim 9$) \citep{Welbanks2024}. In fact, \citet{Welbanks2024} reports difficulty in reproducing this silicate cloud feature with retrieval modeling. Similarly, \citet{Sing2024} finds supporting evidence for an unexpectedly high internal temperature (T$_{\rm int}$ $\sim$ 460 K) and vertical mixing ($\log$${K_{zz}}$$\sim 11$) to explain the lack of observable methane after conducting retrievals on NIRSpec G395H data. Both \citet{Welbanks2024} and \citet{Sing2024} independently found that strong vertical mixing is required to explain these spectral features. Like WASP-39 b, WASP-107 b may also require higher SNR mid-infrared data in order to put stronger constraints on the cloud layer. The host star for this system has a C/O ratio of 0.50$\pm0.1$ and an Mg/Si ratio of 1.08$\pm0.18$.


\begin{figure*}[ht!]
    \centering
    \includegraphics[width=0.75\linewidth]{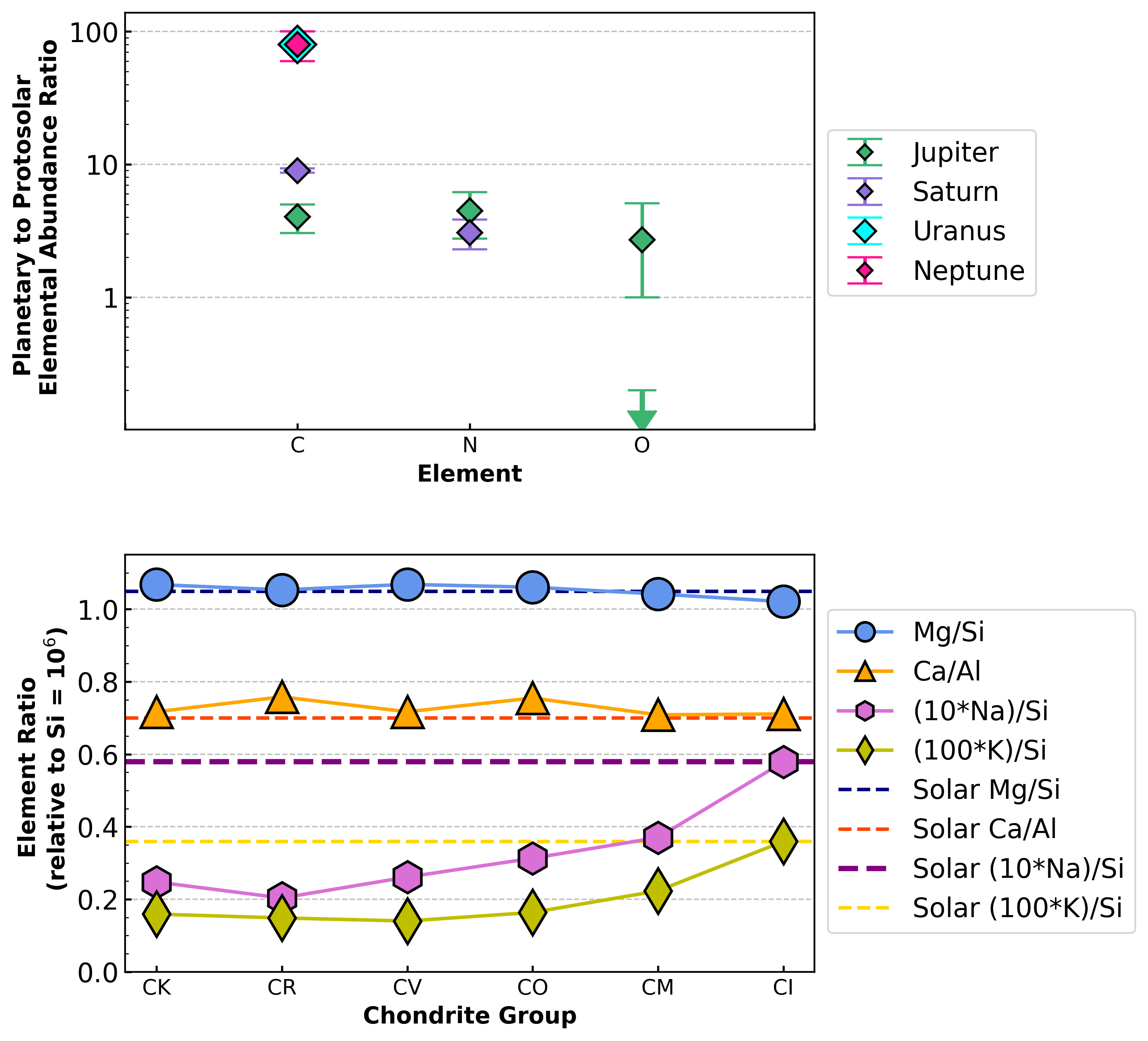}
    \caption{The top figure shows planetary-to-protosolar elemental abundance ratio for major volatile elements (C, N, O) in Jupiter, Saturn, Uranus and Neptune adapted from \citet{Atreya2020}. Protosolar values are based on the solar photospheric values from \citet{Asplund2009}. Measurements of carbon and nitrogen in the giant planets come from a combination of ground and space-based observations as detailed in \citet{Atreya2020} and should be treated with caution due to large uncertainties that make the likelihood of these values being upper or lower limits nonzero. 
    The oxygen measurements for Jupiter come from the Juno Microwave Spectrometer (diamond) \citep{Li2020} and the Galileo Probe Mass Spectrometer (upper limit) \citep{Niemann1996}. The bottom figure shows various element ratios in carbonaceous chondrite groups from \citet{Lodders2021}. Dashed lines represent the respective Solar elemental abundance ratio.}
    \label{fig:chondrites}
\end{figure*}

\section{Host Star Chemistry and Its Implications}\label{sec:host chem}
\subsection{Metallicity and Abundance Ratios}
Relics of planetary formation are found within an object's current atmospheric chemistry. Using bulk chemical abundances to trace back an object's dynamic history is a longstanding approach. Teams have performed analyses across the galaxy \citep{Hawkins2023}, within the local Solar neighborhood \citep{Brewer2016}, as well as among giant planets \citep{Madhusudhan2012, Oberg2011}. For brown dwarfs, science has shown the power, albeit with difficulties, of obtaining bulk chemical abundances \citep[e.g.][]{Calamari2024, Phillips2024}. While challenges persist in determining the bulk C/O ratio, this has still been a continued smoking gun of birth location. More recently, planetary studies have attempted to trace formation location through a bulk volatile-to-refractory ratio, instead \citep[e.g.][]{Lothringer2021, Chachan2023, Pelletier2025}. For companion systems, the accompanying brown dwarf is expected to inherit the same C/O as its host star due to co-evality, contrary to planetary companions. Despite its drawbacks (namely, its potential to be overly simplistic), this has become a powerful metric in substellar modeling as the major molecular absorbers available for study are carbon- and oxygen-based (i.e., H$_2$O, CO, CH$_4$, CO$_2$). So far, a few examples of modeled data support this theorized chemical consistency between brown dwarf companions and their host stars, specifically Gl 570 D, HR 7672 B and Gliese 229 B \citep{Line2015, Wang2022, Calamari2022, Xuan2024}.

Assuming a one-to-one chemical makeup for benchmark companion brown dwarfs places resulting substellar models in much-needed context. \citet{Calamari2024} showed that mapping the known elemental abundances of bright host stars to their associated substellar atmosphere improved our understanding of cloud dynamics and modeling. By accounting for the effects of bulk chemistry and condensation, \citet{Calamari2024} demonstrated that the total oxygen abundance found in gas-phase species (such as H$_2$O, CO, CO$_2$) above known cloud layers was, on average, 17$\%$ lower than the bulk oxygen composition of the atmosphere, providing a correction to estimates of bulk C/O ratios often reported for substellar atmospheres. Additionally, the object's bulk Mg/Si ratio was shown to be the species determinant in known silicate-rich cloud layers. Understanding the larger context within which these co-moving systems formed and evolved therefore aids our model selection and interpretation.

Applications of host star chemistry prove more difficult for objects formed \textit{in-situ} where the available gas-phase and solid-phase chemical inventory of a protoplanetary disk is expected to vary with radial and azimuthal distance \citep[e.g.][and references therein]{Oberg2021}. Substantial work has been done employing ALMA to observationally map the chemical effects of radial disk temperature gradients \citep[see][]{Oberg2023}. Work has focused largely on characterizing CO gas and identifying condensation locations, or ‘‘snowlines", for carbon and oxygen species (mainly, H$_2$O and CO). Radial temperature gradients produce such snowlines as gas transitions to ices in the disk midplane. While the bulk disk chemistry stays consistent on the order of $\sim$10's of au, these ice lines will fractionate the elemental inventory of the disk between the gas and solid phases so that the available material for planetary accretion changes as you move throughout the disk.
\citet{Bosman2021a} reported observations of elevated C/O ratios throughout the disks of three pre-main sequence stars each around C/O$\sim$2, with implications that newly forming gas giants may be accreting envelopes with similarly high C/O ratios (specifically due to substellar [O/H]).

However, connecting protoplanetary disk chemistry to present day atmospheres is a still-emerging task. Disk gas mass places fundamental constraints on what types of planetary systems can form, yet there is no single reliable tracer for it \citep{Oberg2021}. Many common disk volatiles (i.e., H$_2$O, CO$_2$, NH$_3$, N$_2$) that appear in the atmospheres of both hot Jupiters and brown dwarfs are difficult or impossible to observe at millimeter wavelengths, further prohibiting a link between star, disk, and planet chemical inventory. An attempt to bridge this gap is through observations of our own Solar system, where abundances of common volatile or refractory species are observed in the Solar system giants and the meteoritic record.

\subsection{Connecting to Our Solar System Baseline}

Tracing the abundance of volatiles (C, N, O, etc.) in the Solar system giants is a non-trivial task. Several decades of work employing both remote spectroscopy and in-situ measurements have attempted to determine the chemical abundances of the gas giant planets. The most direct measurements come from the Galileo Probe Mass Spectrometer (GPMS) which sampled Jupiter's atmosphere in 1995 \citep{Niemann1992}. In addition to supplementary ground- and space-based observations, dedicated missions to uncover the atmospheres and structures of the giant planets included Voyager \citep{Stone1977, Kohlhase1977}, Cassini \citep{Porco2004} and Juno \citep{Bolton2017}. \citet{Atreya2018, Atreya2020} summarize the breadth of this work in which a key finding was the heavy element enrichment of nearly all measured species across all four planets at two times solar abundance or higher. There are some caveats, namely, the reported depletion of oxygen measured in Jupiter by the GPMS which is thought to be due to a local entry "hot spot" not representative of the global environment.

Follow up observations with the Juno Microwave Spectrometer (MWR) indicate an oxygen enrichment more in line with the other measured volatiles, carbon and nitrogen 
($\sim$1-5.1 times solar) \citep{Li2020}. Figure \ref{fig:chondrites} highlights measured abundances of the volatiles (carbon, nitrogen and oxygen) in the four giant planets. While theory and ALMA observations predict a disk environment depleted in oxygen (elevated C/O ratio relative to solar), analysis from Jupiter suggests a potentially uniform bulk volatile enrichment. This mismatch between observation and theory could point to flaws in our understanding of formation processes, accretion environments, internal thermochemical atmospheric processes and/or evolutionary processes. What remains to be seen is the difficulty in aligning planetary C/O ratio with a disk formation location (i.e., beyond a given snowline) on top of the difficulty in ascertaining bulk volatile abundances for the Solar System giants. 


Alternatively, if we look at the meteoritic record, we find a chemical and mineralogical history that traces back to the earliest formed solids in our Solar system. Of the several known chondritic meteorite classes, carbonaceous chondrite meteorites preserve the first solids to condense in the early Solar System. Unlike achondrites, chondritic meteorites have not undergone differentiation, and therefore represent original Solar system material \citep{Weisberg2006}. Of the major chondrite classes (ordinary, enstatite and carbonaceous), our focus here is on the carbonaceous chondrites. Isotopic differences between carbonaceous and non-carbonaceous chondrites suggest formation in temporally and spatially distinct reservoirs \citep{Warren2011, Kruijer2017}. While there was likely overlap in their formation timelines, carbonaceous chondrites formed slightly earlier (within the first $\sim$1-2 million years) and past the H$_2$O and CO snowlines in the outer disk, alongside the early formation of Jupiter \citep{Kruijer2017}. This isotopic distinction between the non-carbonaceous and carbonaceous chondrites provides insight into the early protoplanetary disk and planet formation models, and suggests that Jupiter likely incorporated refractory material from a resevoir similar to that of the carbonaceous chondrites.

\begin{figure*}[ht!]
    \centering
    \includegraphics[width=0.75\linewidth]{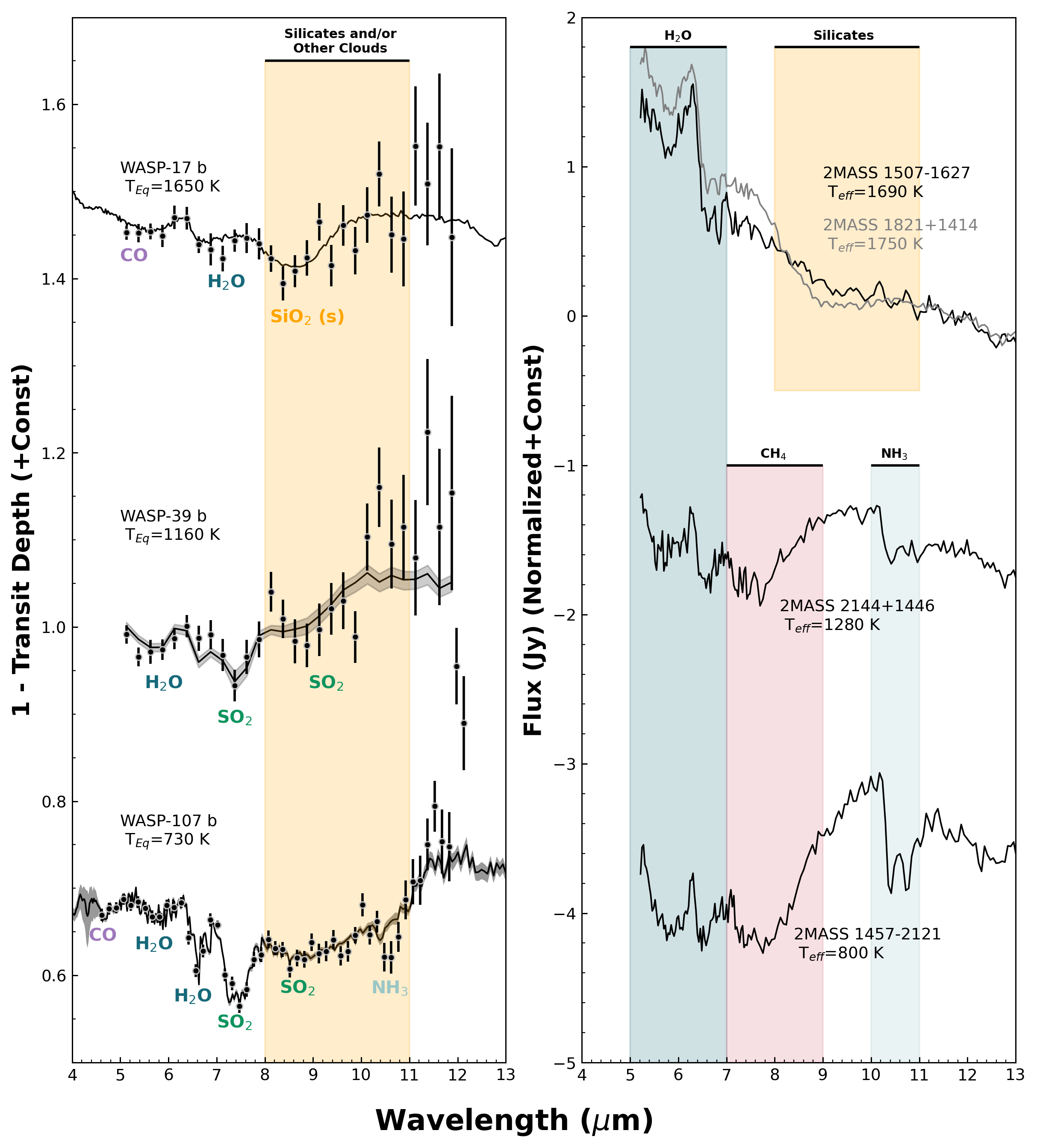}
    \caption{Comparison of mid-infrared spectra of irradiated exoplanet atmospheres and brown dwarfs. In the left panel, we show JWST MIRI transit spectra from $\sim$ 4 - 12 $\mu$m for three giant exoplanets, inverted only as a comparative visual aid to the brown dwarf spectra on the right panel. Observational data (points) and associated best fit retrieval models for the exoplanets are taken from \citet{Grant2023, Powell2024, Dyrek2024} (top to bottom, respectively). Equilibrium temperatures are noted and molecular sources of opacity are labeled. In the right panel, we show MIR brown dwarf spectra from the Spitzer Space telescope, adapted from \citet{Suarez2022}. Effective temperatures and molecular absorption bands are labeled. Typical uncertainties on T$_{\rm eff}$ using \citet{sedkit} are approximately 1-5$\%$.}
    \label{fig:Spectra}
\end{figure*}

Of the eight groups of carbonaceous chondrites (CI, CM, CV, CO, CR, CH, CK, CB), the CI chondrites are considered the closest to solar composition, including volatile element inventory. The remaining groups are notably depleted in volatile content \citep[i.e.,][]{Weisberg2006, Brauk2018, Hellmann2020}. 
Therefore, while not a reference point for protoplanetary disk volatile (C, N, O) content, the carbonaceous chondrites are a standard for non-volatile disk matter in the same reservoir and time frame as the giant planet accretion. Figure \ref{fig:chondrites} highlights elemental abundance ratios for a variety of refractory species (Mg, Si, Ca, Al, Ti) along with the alkali elements (Na, K) in carbonaceous chondrites. The mechanisms that cause volatile elemental abundance variations between chondrite groups, as well as variations from the respective Solar ratio are still heavily debated \citep{Brauk2018, Hellmann2020, vanKooten2024}. However, we note the consistent Mg/Si ratio between carbonaceous chondrite groups as a near match to the Solar value. This suggests that the outer protoplanetary disk maintained a global Mg/Si consistent with Solar. If we expect that Jupiter accreted refractory material from a reservoir similar to that of the carbonaceous chondrites, meteoritic evidence would imply that its bulk Mg/Si ratio should be in line with the Solar value.

\section{Silicate Clouds in Brown Dwarfs and Hot Jupiters}\label{sec:silicates}

Population studies of brown dwarfs have been instrumental in aiding our understanding of atmospheric dynamics, particularly as they relate to cloud formation. While condensing material in brown dwarfs had been theorized and modeled even before the very first observations of Gliese 229 B \citep[i.e.][etc.]{Burrows1993, Tsuji1996, Allard2001, Marley2002}, it was not until observations from the Spitzer Space Telescope nearly ten years later that mid-infrared data confirmed these predictions. \citet{Roellig2004, Cushing2006, Suarez2022} compiled a dataset of field M5 - T9 dwarfs observed by Spitzer (113 objects) to unveil comprehensive observational evidence of silicate clouds as they appear in L dwarfs and subsequently sink below the photosphere at later spectral types. In this section, we summarize the body of this work as it relates to silicate clouds in both brown dwarfs and giant planets.

\subsection{A Color Sequence for the Hot Jupiter Population}

\citet{Suarez2022} presents a re-reduction of all 113 field M5 - T9 dwarfs detected and observed with the Spitzer InfraRed Spectrograph (IRS) (most with 5.2 - 14.2 $\mu$m). In this sample were 12 M5 - M9 dwarfs, 69 L0 - L9 dwarfs and 32 T0 - T9 dwarfs, covering an effective temperature range of approximately 500 - 3,000 K. Spectral sequences of these objects reveal patterns in the strength of molecular and condensate mid-infrared absorbers. The dominant gaseous absorbers in this spectral window vary by spectral type but include H$_2$O (water) at 6.25 $\mu$m, CH$_4$ (methane) at 7.65 $\mu$m, and NH$_3$ (ammonia) at 10.5 $\mu$m. \citet{Cushing2006, Suarez2022} find that water absorption is evident in all spectra and the strength of the feature increases as spectral type increases (i.e., as temperature decreases). Additionally, methane and ammonia features appear consistently by spectral type L8 and T2, respectively, and increase with increasing spectral type. This observational baseline is evidence that supports much of the thermochemical theory that predicts these species under local thermodynamic equilibrium (LTE) conditions (i.e., methane being the dominant carbon-bearing species in atmospheres below approximately 1,300 K) \citep{LodFeg2002}.

For the condensate absorbers, \citet{Suarez2022} comprehensively mapped evidence of silicate clouds in the mid-infrared (8-11 $\mu$m). They find that silicate clouds first become visible in brown dwarf atmospheres around spectral type L2, or at effective temperatures of $\sim$2,000 K. The silicate absorption feature is strongest in mid-L type objects (L4 - L6), or 1,500 K $\leq$ T$_{\rm eff}$ $\leq$ 1,700 K. Past spectral type L8 (T$_{\rm eff}$ = 1,350 K) the detection of the absorption feature wanes and \citet{Suarez2022} speculate that this indicates silicate particles have sunk below the photosphere. Despite lacking a spectral signature, silicate clouds are believed to exist in early T dwarfs at higher pressure levels than probed by the 8 - 11 $\mu$m range (i.e., $\geq$ 1 bar). In fact, photometric variability in T dwarfs is thought to be linked to silicate clouds \citep{Apai2013, Artigau2018}. Additionally, \citet{Suarez2022} note that while L dwarfs are known to be cloudy, not all L-type dwarfs have evidence of silicate absorption in their spectra. A possible explanation of this phenomenon is different viewing orientations, with near pole-on objects exhibiting weak or absent silicate absorption and subsequently bluer colors \citep{Suarez2023}. Finally, objects that do exhibit silicate absorption are redder in the near-infrared than other field dwarfs of the same spectral type \citep{Burgasser2007}.

Compared to work on brown dwarfs, studying giant exoplanets as a population is an emerging task -- one which will require further use of JWST to build out. Unlike the brown dwarf population, where predictive modeling can help guide observation, there is less clarity in what trends should exist for giant exoplanets. A major obstacle in determining population-level atmospheric trends for giant exoplanets is constraining T-P profiles. The available parameter, T$_{\rm eq}$ (the theoretical temperature a planet would have if it were in radiative equilibrium with its host star), is not a totally reliable framework as it neglects the effects atmospheric composition and internal heat can have on its temperature structure \citep[see][]{Marley2008, Madu2019}. Of course, this is complicated even further in tidally locked worlds where the T$_{\rm eq}$ is even less of an accurate metric of the potential thermochemical processes available in an atmosphere. While we can consider utilizing measurements of dayside and nightside temperature for tidally locked planets, T$_{\rm eq}$ is our starting point for these objects. 

Prior to JWST, several works utilized Hubble Space Telescope (HST) spectra or Spitzer photometry of several transiting exoplanets to study trends across a wide range of temperatures (750 $<$ T$_{\rm eq}$ 2600 K). \citet{Triaud2014} presents the first color-magnitude (CMD) diagram for transiting exoplanets using Spitzer IRAC channels 1 and 2 (at 3.6 and 4.5 $\mu$m, respectively). For the eight objects studied, they find that observations of the dayside emission -- which lie in M and L dwarf temperature space -- broadly follow the brown dwarf CMD. Then, \citet{Sing2016} published a spectral sequence of ten transiting exoplanets covering 0.3 - 5.0 $\mu$m from combined Spitzer and HST observations. They order transmission spectra by $\Delta$Z$_{UB-LM}$ (a spectral index used to compare the relative strength of scattering in the 0.3 - 0.57 $\mu$m region to that of molecular absorption in the 3 - 5 $\mu$m region) and find a transition from clear to cloudy objects. Additionally, \citet{Gao2020} showed that observed trends in the 1.4$\mu$m water feature strength from HST transit spectra are reproduced by predictive cloud models across a range of T$_{\rm eq}$ -- specifically, that aerosol opacity is dominated by silicate clouds in atmospheres with T$_{\rm eq}$ $>$ 950 K. More recently, \citet{Fu2025} attempted to uncover population-level trends on the currently available JWST near-infrared (2.7 - 5 $\mu$m) transit data by looking at correlations between stellar and planetary temperature, planetary mass and surface gravity. They define four wavelength bands covering major carbon, nitrogen, oxygen and sulfur bearing absorbers similar to photometric color for brown dwarfs \citep[see][]{Kirkpatrick1991, Kirkpatrick1999}. While they suggest there may be some relationship between these fundamental and spectral parameters, they, too, require additional data from JWST.

At present, we can compare the JWST MIRI transit data of WASP-17 b, WASP-39 b and WASP-107 b not only against each other but also against Spitzer IRS data of brown dwarfs of similar temperature. Again, we do acknowledge the inability to use T$_{\rm eq}$ and T$_{\rm eff}$ as 1:1 comparisons for what we would expect thermochemically in these atmospheres. However, we use this as a starting point. In \autoref{fig:Spectra}, we compare the JWST data and best fit retrieval models as presented in \citet{Grant2023, Powell2024, Dyrek2024} against four brown dwarf spectra from \citet{Suarez2022}. The chosen brown dwarf spectra with corresponding temperatures are 2MASS 1507-1627, an L5 dwarf with no spectral signature of silicates, 2MASS 1821+1414, an L5 dwarf with a strong spectral signature of silicates, 2MASS 2144+1446, a T2.5 dwarf, and 2MASS 1457-2121, a T7 dwarf. Diverging from convention, we plot the inverse of transit depth to give a visual aid for comparisons to the brown dwarf spectral flux in the right-hand panel. Molecular absorption bands for the brown dwarf sequence are labeled as they were in \citet{Suarez2022} and T$_{\rm eff}$ have been calculated from SED-fitting using the Spitzer spectra following \citet{Filippazzo2015, sedkit}. While we cannot yet label absorption bands across a sequence for the giant exoplanet sample, we do label relevant opacity features as they were presented in their respective retrieval papers.

Similar to the brown dwarf population, we see the presence of water in all three exoplanet atmospheres. For WASP-17 b and WASP-107 b, we can see the relative strength of this water feature at 6.25 $\mu$m increase, as expected, with decreasing temperature. In WASP-39 b, water is only detected slightly blue-ward of 6.25 $\mu$m. Additionally, we see evidence for ammonia in the coolest of the three objects, WASP-107 b which is in line with thermochemical predictions in this temperature regime and matches well with the mirrored T dwarf absorption. No methane is detected in any of these three exoplanets which is contrary to what we might expect, especially for WASP-107 b. It is possible that SO$_2$, a strong photochemically created molecular opacity source for irradiated atmospheres, obscures the 7.65 $\mu$m feature. However, lack of methane detections in exoplanet atmospheres is also attributed to chemical quench processes \citep{Fortney2020, Sing2024} and/or photochemical dissociation \citep{Hobbs2021}. Finally, we can see in WASP-17 b a clear opacity source beginning at 8 $\mu$m, where silicate absorption begins. In the brown dwarf panel, this flux depression due to silicate absorption is evident between the two L dwarfs, 2MASS 1821+1414 (cloudy) and 2MASS 1507-1627 (cloudless). We can tentatively see the difference in opacity at this 8 $\mu$m region between WASP-17 b, where quartz clouds were definitively detected, and WASP-39 b and WASP-107 b, where models struggled to constrain cloud properties. However, it is likely that higher resolution and SNR mid-infrared data using JWST's MIRI MRS instrument mode will be needed to understand these trends further.

\begin{figure}
   \centering
   \includegraphics[width=\linewidth]{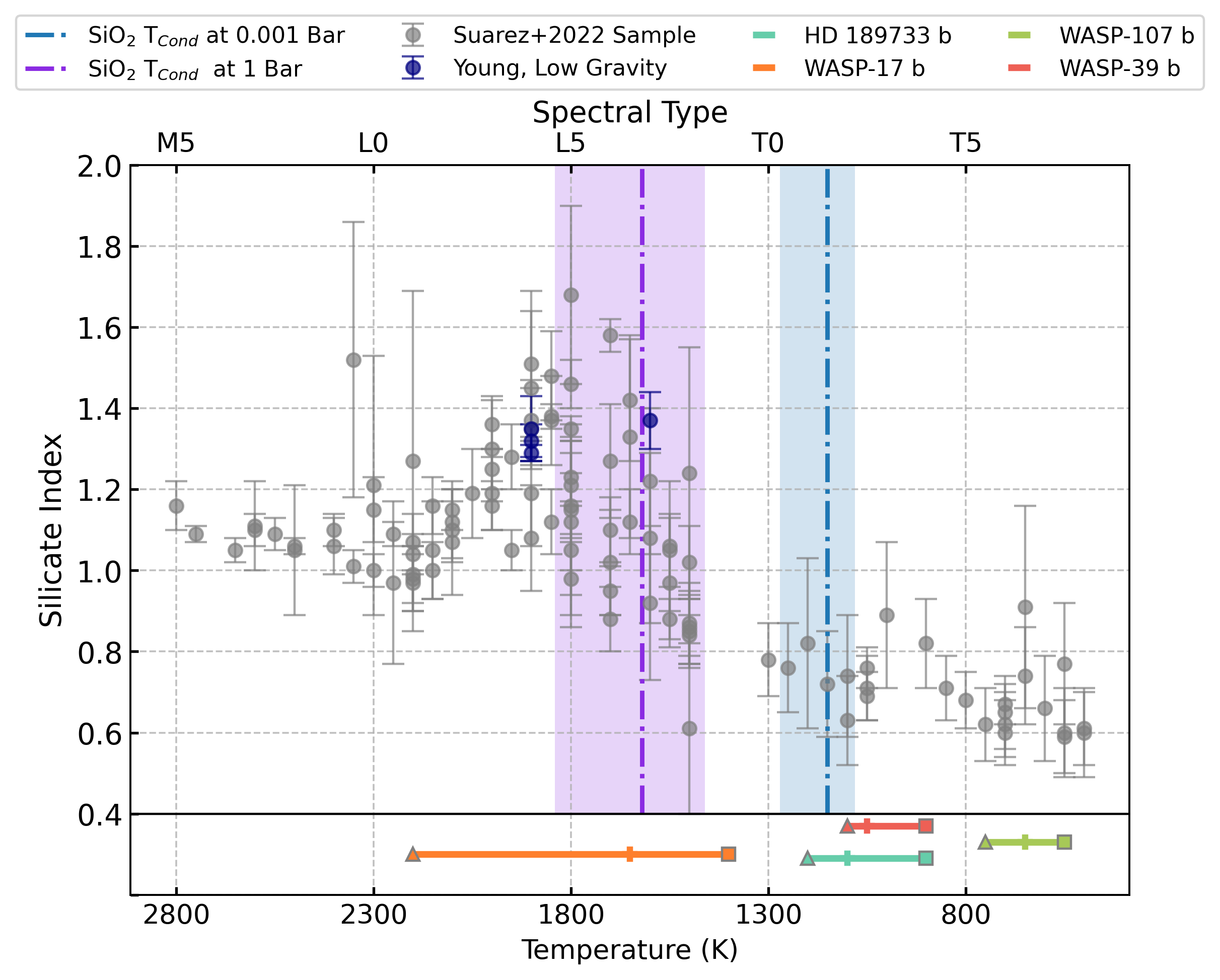}
   \caption{We show an adaptation of silicate index versus spectral type (or, temperature) as published in \citet{Suarez2022}. Silicate index for the brown dwarf population observed with the Spitzer Space Telescope (grey points) is discussed in \citet{Suarez2022}. Additionally, we highlight in navy four brown dwarfs categorized as young, low gravity objects from \citet{Suarez2023}. At the bottom, we plot the range of temperatures for the four JWST-observed and retrieved giant exoplanets explored in this work, indicating where they might land on this distribution if a silicate index were possible to calculate. Dayside temperature is indicated by a triangle, nightside temperature by a square, and T$_{\rm eq}$ by an intersecting line. We also plot the condensation temperatures for quartz at 1 bar (purple) and 0.001 bar (blue) as estimates of photospheric silicate condensation points in brown dwarfs and giant planets, respectively. Corresponding colored bands represent the range of T$_{\rm Cond}$ for quartz at [Fe/H]=0.1xSolar (cooler) and [Fe/H]=10xSolar (warmer).}
   \label{fig:SilInd}
\end{figure}

\subsection{Onset of Silicate Absorption via Silicate Index}

The strength of the 8 $\mu$m silicate absorption feature in brown dwarf spectra was quantified in \citet{Suarez2022} via the \textit{silicate index}. They define this parameter as the ratio of the interpolated continuum flux at 9 $\mu$m to the average flux in a wavelength interval of 0.6 $\mu$m centered at 9 $\mu$m. This effectively gives a numerical designation to the strength of this feature such that increased silicate index corresponds to increased silicate absorption strength. As previously mentioned, they find a peak in the silicate index around spectral types L4-L6, with discernible silicate absorption beginning at spectral type L2 and ending at spectral type L8.

\begin{figure*}[ht]
    \centering
    \includegraphics[width=0.7\linewidth]{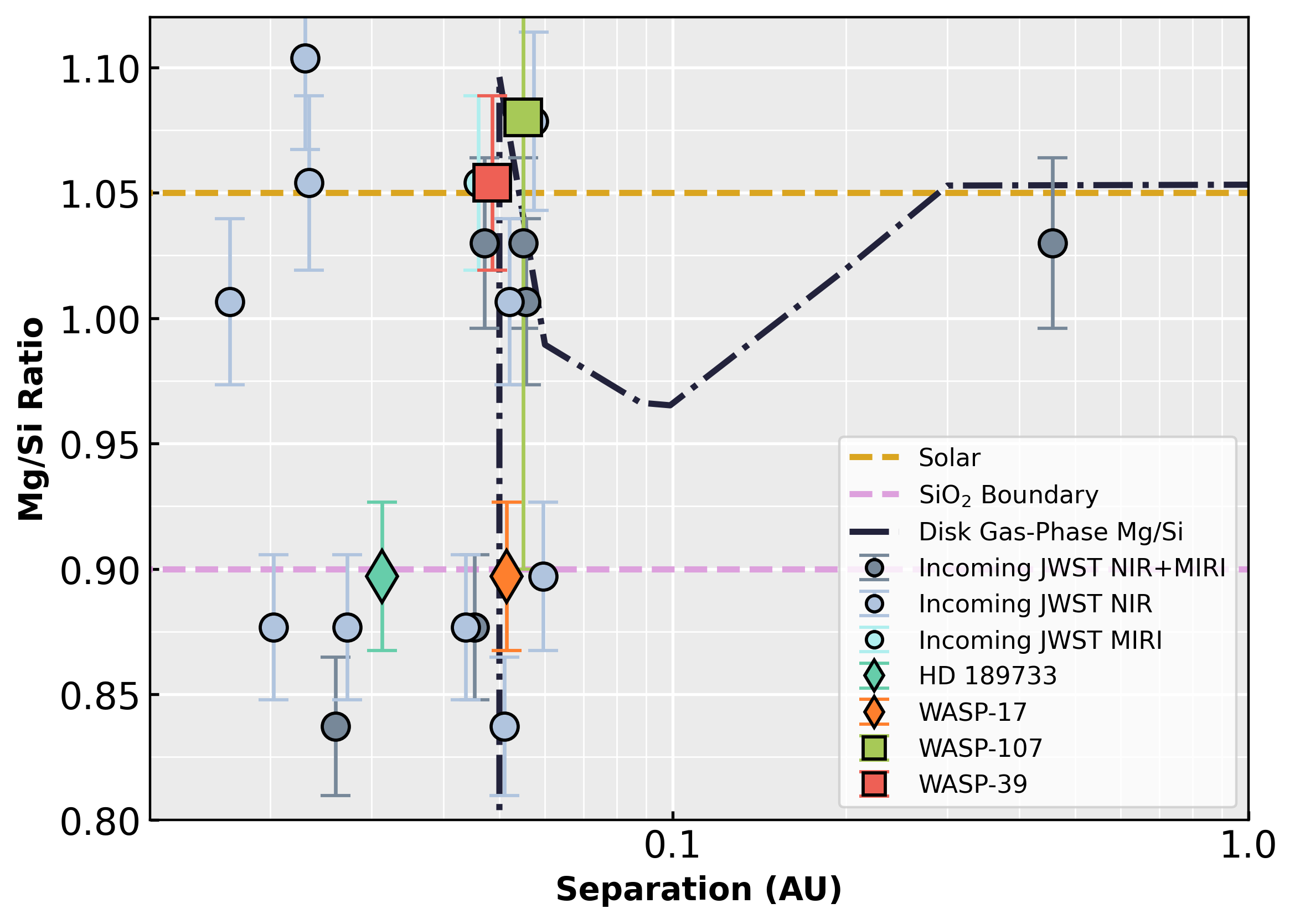}
    \caption{Mg/Si ratios of JWST-observed hot Jupiter host stars versus separation (semi-major axis). Mg/Si ratios compiled from \citet{Brewer2016} with an error of 3.3$\%$ as quoted in \citet{Brewer2016b} as well as \citet{Hejazi2023}. Diamond markers indicate systems in which the hot Jupiter companion is reported to have spectral signatures of SiO$_2$(s) clouds. Square markers indicate no current robust detection for a particular silicate species. The purple dashed line indicates the Mg/Si ratio point that distinguishes the onset of either forsterite ($>$0.9) or quartz ($<$0.9) condensates \citep{Calamari2024}. The navy dashed line indicates the theoretical gas-phase Mg/Si ratio in a solar-type disk from \citet{Thiabaud2015}.}
    \label{fig:MgSiDisk}
\end{figure*}

In \autoref{fig:SilInd}, we adapt Figure 6 from \citet{Suarez2022} showing silicate index versus spectral type (or, temperature). We include vertical lines indicating the 50$\%$ condensation point of quartz at 1 bar and 0.001 bar in a Solar composition gas (i.e. the point in T-P space at which 50$\%$ of silicon gas would have condensed out) \citep{Lodders2003, Visscher2010} to represent roughly the observable photospheric pressure in brown dwarfs and giant planets, respectively. We included a shaded region around each vertical line to show the range of T$_{\rm Cond}$ under differing bulk metallicities; at [Fe/H]=0.1xSolar, quartz condenses at lower temperatures and at [Fe/H]=10xSolar, quartz condense at higher temperatures. We can see that the condensation point for quartz at 1 bar is close to the peak of the silicate index distribution. Additionally, below the figure showing silicate indices for the \citet{Suarez2022} Spitzer sample, we plot the four exoplanets discussed in this work. As mentioned, because T$_{\rm eq}$ is not comparable to T$_{\rm eff}$, we plot these as distributions ranging from dayside to nightside temperature, noting stated T$_{\rm eq}$, as well. We do this in order to capture the range of possible temperatures where silicate condensation chemistry may be visible. Of course, calculating a silicate index for this population would not be as straightforward as there is no discrete continuum from which to compare opacity strength as in the brown dwarf case.

Of the four exoplanets, WASP-17 b is the only object which coincides with silicate condensation for brown dwarfs. At T$_{\rm eq}$=1650 K this lies exactly within the expected temperature range where silicate absorption peaks. For HD 189733 b, WASP-39 b and WASP-107 b, we can see that their temperature ranges sit entirely within T dwarf temperatures, bringing into question the feasible observability of silicate clouds. However, these objects do have relatively lower gravities which works in favor of silicate cloud visibility. From the brown dwarf population, we have evidence that sedimentation efficiency is lower in young, low gravity objects, allowing clouds to remain visible at lower temperatures \citet{Suarez2023}. We have included four young, low gravity brown dwarfs in \autoref{fig:SilInd} as a closer comparison to the giant exoplanet population. However, more objects will be needed to understand changes in silicate index trend from field age objects. Additionally, as we expect silicate cloud layers to exist deeper in T dwarf atmospheres, it is also possible that we can see residual spectral effects of this, akin to photometric variability, in the exoplanet population as well. Alternatively, the 50$\%$ condensation point of SiO$_2$ at 0.001 bar begins roughly around spectral type T2 or, 1,100 K. Although silicate features disappear by this temperature in brown dwarfs, observational evidence, like in the case of HD 189733 b, shows that silicates may still be visible for exoplanets in this temperature regime. While we do not have a comparable method by which to calculate silicate index for the exoplanet population, we might imagine this silicate index trend to peak at later temperatures. We, again, await further JWST data and cloud modeling results in order to draw robust comparisons to the brown dwarf population.

\section{Joining Theory, Models and Data}\label{sec:clouds}
\subsection{Current Cloud Model Agreement}

\begin{figure*}[ht!]
   \centering
   \includegraphics[width=0.75\linewidth]{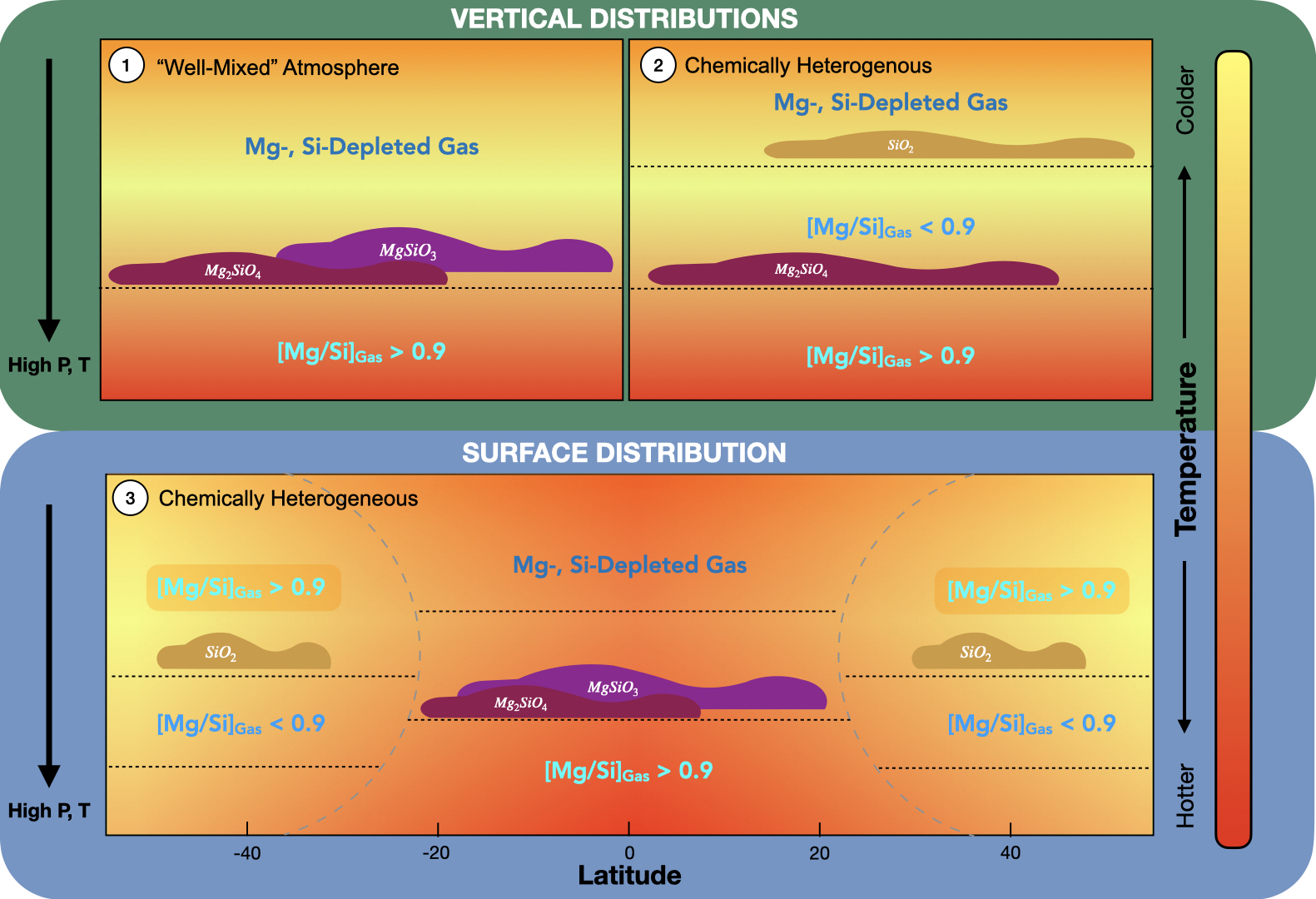}
   \caption{Schematic detailing potential atmospheric distributions of silicate cloud species with altitude. Vertical distributions (top panel) show 1-D slices of an atmosphere, descending from top (low pressure, temperature) to bottom (high pressure, temperature). On the top left is the case for a “well-mixed" atmosphere where pressure layers are not chemically differentiated. On the top right is a case for a chemically inhomogenous atmosphere where convective cycling may inhibit atmospheric pressure layers from equilibrating. A surface distribution (bottom panel) shows 1-D slices of an atmosphere along different latitudinal regions where chemical inhomogeneity may not only be due to vertical mixing but also driven by differences in horizontal transport from equator to pole. Black dashed lines represent condensation points for different silicate species (not to scale). Background color correlates to atmospheric temperatures of irradiated atmospheres, although these types of dynamics may also be present in non-irradiated atmospheres. All three scenarios assume a given object has a bulk Mg/Si ratio $>$ 0.9.}
   \label{fig:schema}
\end{figure*}

While bulk volatile element inventory remains challenging to trace in the protoplanetary disk, meteorites and present-day planetary atmospheres, a through-line may be in the refractory element ratio of Mg/Si. Although challenging to measure directly from brown dwarf or exoplanet spectra, the bulk Mg/Si ratio is an important arbiter of silicate cloud species theorized and observed in cool atmospheres \citep[see][]{Suarez2022, Visscher2010, Calamari2024}. For benchmark brown dwarf systems where the host star chemistry is more readily available, a bulk Mg/Si ratio can be determined for the system and used in subsequent analysis \citep{Calamari2024}. For systems where a companion formed within the protoplanetary disk, assuming total chemical consistency is not possible. However, we turn back to our understanding of the chemical distribution of our own primordial disk material through meteoritic data. From Figure \ref{fig:chondrites}, we can see that the primitive Mg/Si ratio in the outer disk as traced by the carbonaceous chondrites is not only constant between chondrite group but also is a near match to the solar value. Additionally, in Figure \ref{fig:MgSiDisk} we can see the theoretical prediction of how gaseous Mg/Si ratio changes with radial distance from the star \citep{Thiabaud2015}. Beyond 0.3 au for a solar-type disk, Mg/Si ratio is predicted to level out to the stellar value. Therefore, if we understand giant gaseous planets to be forming and accreting material largely from the outer disk \citep{Lunine2007}, we can expect these planets to adopt the same Mg/Si ratio as their host star.

According to the stoichiometric analysis in \citet{Calamari2024}, for systems in which the temperatures and pressures lend themselves to silicate cloud condensation \citep[see][]{Visscher2010}, if the Mg/Si ratio is $<$ 0.9 then enstatite and quartz clouds will form. If the Mg/Si ratio is $>$ 0.9, enstatite and forsterite clouds will condense. If giant exoplanets inherit their host star Mg/Si ratio, we should be able to predict their dominant silicate cloud species under equilibrium conditions. Figure \ref{fig:MgSiDisk} highlights the four JWST-observed giant exoplanet systems for which retrieval analysis has determined a best-fitting model that includes clouds. For the two systems where the host star Mg/Si ratio is $<$ 0.9, studies reported spectral and statistical evidence for SiO$_2$(s) clouds \citep{Grant2023, Inglis2024}, as we might expect. It is also relevant to note that recent studies of JWST-observed ultra-hot Jupiters are finding statistical consistency in the gas-phase Mg/Si ratios between star and planet, as in the case of WASP-189 b \citep{Sanchez2025}. In the two systems where the host Mg/Si ratio is $>$ 0.9, determinations of cloud species is not quite clear yet. Although both WASP-107 b and WASP-39 b have predictions for enstatite being the dominant silicate absorber, neither study was able to put robust statistical or theoretical constraints on cloud parameters (see \autoref{sec:systems}). \citet{Dyrek2024} does claim evidence of silicate clouds in the upper atmosphere of WASP-107 b. However, not only would we not expect quartz to form based on its host Mg/Si ratio, but discussions of its atmosphere being too cool to condense silicate clouds at visible pressure layers also puts this into question. We discuss the likelihood of observable silicate clouds at cooler temperatures in \autoref{sec:silicates}.


Of course, we know from our our Solar system giants the complexity of these differentiated atmospheres. Studies from Jupiter have revealed the complicated disequilibrium chemistry driven largely by convective mixing and turbulent storms \citep[i.e.][]{Prinn1977, visscher2010icarus, Wang2015, Visscher2020, Moeckel2025}. This type of variegated vertical and horizontal structure is likely even more severe in the case of tidally locked systems \citep[i.e.][]{Menou2009, Parmentier2013, Showman2015, Komacek2016}. Therefore, it may be the case that while these giant planets may inherit their host Mg/Si ratio at formation, atmospheric dynamics may drive chemically distinct pressure layers. For example, while an atmosphere may have a \textit{bulk} Mg/Si ratio $>$ 0.9, a local magnesium depletion may trigger the formation of quartz clouds in place of Mg-silicates. Future cloud modeling work will be needed to further investigate the inherited versus local chemistry and thermodynamics of these types of atmospheres.

\begin{figure}[t!]
    \includegraphics[width=\linewidth]{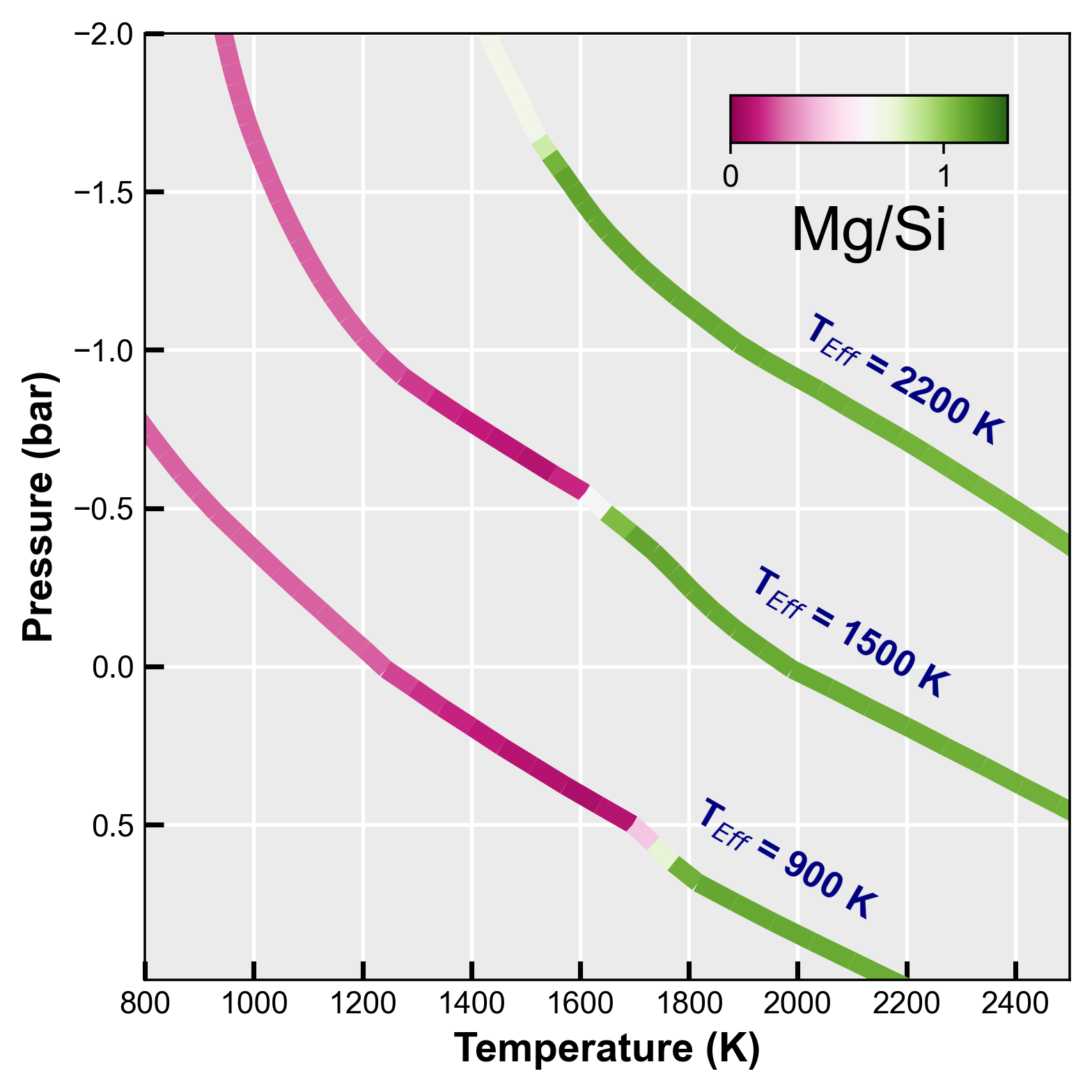}
    \caption{This figure details how gaseous Mg/Si ratio changes as a function of pressure and temperature under thermochemical equilibrium in substellar mass atmospheres. Three P-T curves are shown for T$_{\rm eff}$ = 2,200, 1,500, 900 K from \citet{Morley2024}, uniformly chosen at solar metallicity and C/O and $f_{sed}$=4, log($g$)=4.0. Green P-T points indicate a gaseous bulk Mg/Si = 1.05 whereas pink P-T points indicate regions where refractory material has condensed out, driving gaseous Mg/Si to zero.
    }
    \label{fig:MgSiAlt}
\end{figure}

\subsection{Predictions for Future JWST Retrievals}

In order to better understand whether we can see the chemical fingerprints of planetary formation through silicate cloud modeling, the sample size of retrieved giant exoplanets must increase. Based on accepted programs from JWST Cycles 1 - 4, there are 17 additional giant exoplanets (with masses 0.25 $<$ M$_{Jup}$ $<$ 13) with measured host star abundances whose spectroscopic data will serve to test this prediction (see Figure \ref{fig:MgSiDisk}). Of the 17 planets, only six will have both incoming or archived near-infrared and mid-infrared spectral coverage, which is likely needed in order to robustly detect and characterize silicate condensation.

The aforementioned giant exoplanets not yet discussed in this work are HD 209458 b (GO 2667 PI: Wakeford), WASP-121 b (GO 2961 PI: Molliere), WASP-69 b (GO 3712 PI: Cubillos, GTO PI: Greene), HD 80606 b (GO 2008 PI: Kataria, GO 2488 PI: Sikora), 
HAT-P-1 b (GO 2950 PI: Waters), Kepler-12 b (GO 7849 PI: Moran). 
Based on their host Mg/Si ratios and T$_{\rm eq}$, we might expect to see quartz clouds in the atmospheres of WASP-121 b and WASP-69 b. Recently, \citet{EvansSoma2025} conducted retrieval modeling on near-infrared (2.73–5.17 $\mu$m) spectroscopic data from JWST for WASP-121 b but could not give definitive evidence for the presence of silicate clouds. Modeling attempts on an extended dataset will likely be needed to characterize clouds in this atmosphere. For HD 209458 b, HD 80606 b, HAT-P-1 b and Kepler-12 b, the Mg/Si ratios of their host stars (all $>$ 0.9) suggest that the formation of quartz clouds would be thermochemically disfavored if these planets do, in fact, inherit their host Mg/Si ratio. Instead, we might expect to see the detection of enstatite and/or forsterite clouds in these planets.

However, while we might expect these condensates in thermodynamically and chemically “well-mixed" atmospheres (i.e., a more ideal convective atmosphere), we do not discount the increased complexity of irradiated and tidally locked atmospheres. Even in the case of isolated brown dwarf atmospheres that can be taken as a first order approximation for the giant exoplanet population, we have observational evidence of chemical disequilibrium due to strong vertical mixing that can chemically striate pressure layers. For example, \citet{Faherty2025} discovered the first observational evidence of silane (SiH$_4$) in the photosphere of a cold world, WISEA J153429.75-104303.3 (WISE 1534), (T$_{\rm eff}$ $<$ 500 K) at an abundance much greater than that predicted by chemical equilibrium. Theory predicts the thermochemical stability of silane in the upper atmosphere as a result of convective timescales being greater than chemical timescales (or, the time it takes for the atmosphere to equilibrate). This type of chemical disequilibrium is not uncommon and has been known to exist even in hotter atmospheres when looking at CO and CO$_{2}$ abundances \citep[e.g.][]{Oppenheimer1998, Stephens2009, Miles2020}.

For the giant exoplanet population, contradictory cloud model results would further highlight dynamics that could create chemically distinct atmospheric pressure layers (i.e., scenarios where the Mg/Si ratio varies significantly with altitude). Figure \ref{fig:schema} presents a simplistic theoretical schema for atmospheres where a bulk inherited Mg/Si ratio would result in the formation of enstatite and forsterite clouds but atmospheric dynamics create observable quartz clouds. We show three potential scenarios for an object with a bulk Mg/Si ratio $>$ 0.9. In one case, we have an idealized, “well-mixed" atmosphere where the hotter, deeper pressure layers are reflective of the bulk chemistry that condenses out enstatite and forsterite as gas rises and cools. Above this cloud layer is an atmosphere depleted in magnesium and silicon, as a majority, if not all, of the refractive species have been locked into clouds.


Another case depicts a chemically heterogeneous scenario where the vertical atmosphere is thought to have pressure layers that are chemically distinct and are separated by convective dynamics in such a way that they do not communicate. As gas rises from hotter pressure layers, forsterite condensation occurs first (roughly 1,800 K at 1 bar) before enstatite condensation (roughly 1,700 K at 1 bar) \citep{Visscher2010}. It is at this narrow P-T point in the atmosphere where we have silicon-enriched gas, prior to enstatite condensation which would trigger full depletion of silicon from the gas-phase. We demonstrate this in \autoref{fig:MgSiAlt}.

In \autoref{fig:MgSiAlt}, we map the Mg/Si ratio as it changes with pressure and temperature in three substellar mass atmospheres. We show P-T profiles from \citet{Morley2024} with Solar metallicity and C/O ratio, $f_{sed}$=4, log($g$)=4.0, and varying effective temperatures (T$_{\rm eff}$=2,200, 1,500, 900 K). In the top profile (T$_{\rm eff}$=2,200 K), we can see that this atmosphere remains too hot for full silicate condensation. In the cooler profiles (T$_{\rm eff}$=1,500 and 900 K), we can see the point in each atmosphere where silicate condensation begins. As forsterite condensation preferentially sequesters twice the amount of magnesium atoms per oxygen than it does for silicon, there is a narrow P-T region where the Mg/Si ratio falls below 0.9 before complete gaseous molecular depletion of the refractory species. If these pressure layers with distinctly varied Mg/Si ratio (i.e., $<$0.9 and $>$0.9) remain disconnected, we can conceivably imagine a P-T region where quartz is thermodynamically favored.

If we return to the heterogenous case in \autoref{fig:schema}, we depict the thermochemistry of \autoref{fig:MgSiAlt} graphically: a detached, cooler pressure layer adopts an Mg/Si ratio $<$ 0.9 and the ability to condense quartz clouds. Again, for this to be thermochemically stable, these regions of the atmosphere must remain dynamically separated as full-scale mixing would cause the destruction of quartz.

Finally, we pose a scenario where chemical inhomogeneities can exist spatially from equator to pole. Due to equatorial jets, differing longitudinal rotation rates or day-to-nightside heat redistribution specifically in tidally locked planets, we may observe atmospheres where the bulk Mg/Si $>$ 0.9 (perhaps reflected in equatorial regions where temperatures are hotter and convective mixing is faster), but the polar regions are chemically detached with Mg/Si ratios $<$ 0.9. Of course, we can imagine atmospheres with chemically heterogeneous vertical and surface distributions, complicating accurate cloud modeling even further. While we model this schematic with irradiated objects in mind, we also acknowledge the possibility for these types of vertical and horizontal transport processes to exist in isolated brown dwarf atmospheres, since we already have evidence of disequilibrium dynamics in several observed worlds. In order to understand the types of chemistry that are inherited and/or the dynamics that cause certain cloud species to appear, more retrieval work on giant exoplanets is needed to obtain a holistic view on this population of objects.

\section{Conclusions} \label{sec:conclusion}

In this work, we explore the ways in which advancements in our understanding of brown dwarf science can help inform modeling and interpretations of giant exoplanet atmospheres. We examine the retrieval cloud modeling results of WASP-17 b, HD 189733 b, WASP-39 b and WASP-107 b as these transiting hot Jupiters have all been observed with JWST MIRI LRS. \citet{Calamari2024} showed the advantages of utilizing host star chemistry to inform brown dwarf modeling in companion systems. As these four exoplanets orbit F, G, and K type stars with available elemental abundances, we examine how host star chemistry can inform the atmospheric modeling and formation history of these objects akin to the brown dwarf population.

In \citet{Calamari2024}, we were able to map the detailed chemistry of the host star onto its substellar companion as a result of assumed coevality. We found that bulk Mg/Si determines the types of silicate species condensing in substellar atmospheres and as such, we were able to place predictions on the types of clouds we might see given a particular host Mg/Si ratio. While this 1:1 elemental mapping is not possible for the giant exoplanet population, we find that the two systems that have conclusive evidence for SiO$_2$(s) clouds -- WASP-17 b and HD 189733 b -- both have host stars with Mg/Si ratios $\leq$ 0.9, the point below which we would expect enstatite and/or quartz clouds to form. Additionally, we ground our work in what we know from our own Solar system by looking at the meteoritic record. Carbonaceous chondrites, objects that preserve the chemistry of the first solid material to condense in the protoplanetary disk, also exhibit a Solar Mg/Si ratio. From theory and observation, we predict that, while the bulk chemical abundances of stars and their planets are likely not homogeneous, giant gaseous exoplanets inherit the same bulk Mg/Si ratio as that of their host star. Further cloud modeling on JWST MIRI transit and/or emission spectra will be needed to illuminate whether we can see the impact of host star chemistry on exoplanet atmospheres or if complex atmospheric dynamics obscure the ideal case of viewing bulk chemistry.

Finally, we compare the JWST MIRI LRS transit data of WASP-17 b, WASP-39 b and WASP-107 b against Spitzer IRS data of field brown dwarfs of related temperature. While it is nontrivial to establish a temperature sequence for irradiated and tidally locked atmospheres, we can learn about the types of expected chemistry through comparisons of hotter to cooler atmospheres. Mirroring the brown dwarf population, we see prevalent water opacity that increases as temperature decreases as well as ammonia opacity in the coolest object, WASP-107 b. We also see the strength of silicate opacity in WASP-17 b, an object that falls closer to L dwarf temperatures, whereas WASP-39 b and WASP-107 b, objects closer to the T dwarf temperature regime, do not show the same spectral indicator. This follows convention for brown dwarfs where silicate absorption is spectrally differentiable in L dwarfs but not in T dwarfs as the cloud layer is predicted to fall below the visible photosphere to deeper pressure layers.

The task of deciphering the complex cloud species and dynamics in exoplanet atmospheres has just begun to unfold given the wealth and quality of data from JWST. In order to best inform atmospheric modeling, we may try to apply the lessons learned from the brown dwarf population, despite the added complexities in these irradiated worlds. Subsequent modeling on incoming mid-infrared datasets as well as additional JWST programs to obtain MIRI spectral coverage of these objects will be instrumental in building out a sample of giant exoplanets for which we can conduct population-wide studies.

\section*{Acknowledgments}
This work was supported by the National Science Foundation via awards AST-2503537, AST-1909776 and AST-1909837 and NASA via award 80NSSC22K0142.

\bibliography{refs, newrefs}
\bibliographystyle{aasjournal}


\end{document}